\documentclass[aip,apl,groupedaddress,reprint,floatfix,nobalancelastpage,amsmath,amsfonts,amssymb]{revtex4-1}
\usepackage{graphicx}

\begin{document}

  \title{Off-resonant coupling between a single quantum dot and a nanobeam photonic crystal cavity}

  \author{Armand Rundquist}
  \email{armandhr@stanford.edu}
  \author{Arka Majumdar}
  \author{Jelena Vu\v{c}kovi\'{c}}
  \affiliation{E. L. Ginzton Laboratory, Stanford University, Stanford, California 94305, USA}
  \date{\today}

  \begin{abstract}
    We demonstrate off-resonant coupling between a single quantum dot and a nanobeam photonic crystal cavity, under resonant excitation of the quantum dot or the cavity. These results are consistent with previous descriptions of off-resonant coupling as an incoherent phonon-mediated process. The extension of this phenomenon to a nanobeam photonic crystal cavity presents interesting possibilities for coherent control of this interaction by tailoring the phonon density of states.
  \end{abstract}


  \maketitle

  One of the most promising platforms for solid state cavity quantum electrodynamics (CQED) is provided by a semiconductor quantum dot (QD) coupled to a photonic crystal cavity \cite{article:eng07, andrei_njp}. Although the initial experiments with this system were primarily motivated by its atomic counterpart, constant interaction of the QD with its fluctuating environment gives rise to several novel phenomena, specific to this solid state system. One of these newly observed phenomena is the off-resonant dot-cavity coupling \cite{article:michler09, article:majumdar09}. Under resonant excitation of the QD, this off-resonant coupling is incoherent, phonon-mediated \cite{majumdar_phonon_11} and has recently received considerable attention because of its potential application in performing resonant QD spectroscopy \cite{majumdar_QD_splitting}, and probing the giant optical Stark shift \cite{waks_starkshift_APL2011}. While this effect has already been demonstrated and thoroughly modeled in a two-dimensional photonic crystal slab cavity, it is worthwhile to explore additional cavity geometries whose features may provide better opportunities to coherently control and enhance the off-resonant interaction. In particular, nanobeam photonic crystal cavities have shown great promise for their small footprint and mode volume ($V_m$), high quality factor \cite{taniyama_ultrahighQnanobeam_OpEx2008, loncar_highQnanobeam_APL2009, notomi_ultrahighQnanobeam_OpEx2010} ($Q$), ease of coupling to on-chip waveguides \cite{loncar_nanobeam2waveguide_APL2010}, and good optomechanical properties \cite{painter_OMcrystals_Nature2009}. The recent demonstration of strong coupling in a nanobeam photonic crystal further emphasizes their potential as a practical implementation of a quantum dot CQED system \cite{arakawa_strongcoupling_APL2011}. Since off-resonant coupling fundamentally relies on the phonon modes available, nanobeam cavities may provide a way to manipulate this effect through control of their mechanical properties. The coherent control of the QD-phonon interaction opens up several avenues in fundamental solid state CQED research, for example, cooling the resonator to ground state \cite{imamoglu_lasercooling_PRL2004} or generation of indistinguishable single photons on demand (by reducing jitter time resulting from phonon assisted relaxation between quantum dot levels) \cite{vuckovic_nonclassical_PhysicaE2006}. In this paper, we demonstrate phonon-mediated off-resonant coupling between a QD and a nanobeam photonic crystal cavity, under resonant excitation of either the QD or the cavity.
  
  \begin{figure}
    \includegraphics{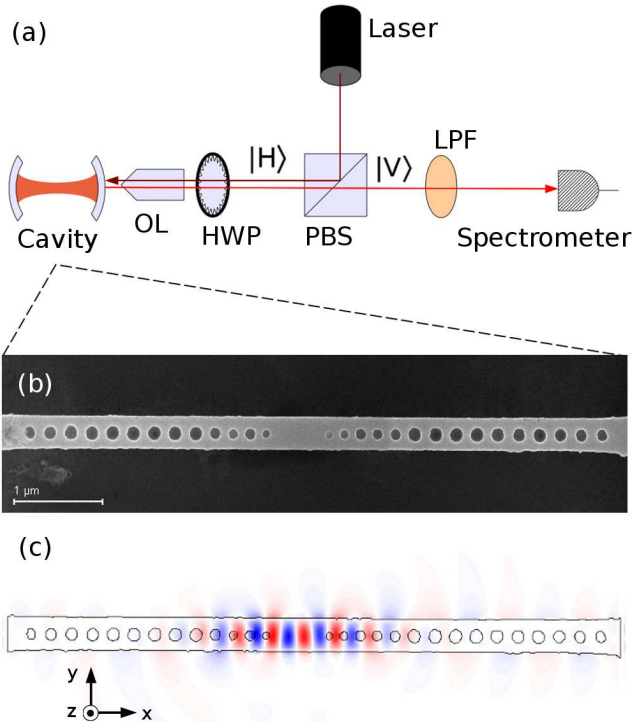}
    \caption{(Color online) (a) The cross-polarized reflectivity setup used to probe the cavity. The excitation laser is passed through a polarizing beam splitter (PBS) and half-wave plate (HWP) before being focused by an objective lens (OL) onto the sample, which is kept under vacuum at a temperature of 30K. Light is collected from the sample by the OL and passes through the HWP and PBS again, which select for the opposite polarization; only light emitted by the sample itself and not directly reflected will have a component in this polarization. A long-pass filter (LPF) can be used to eliminate the excitation source if necessary. (b) SEM of the nanobeam photonic crystal. (c) The $E_y$ field profile of the cavity mode of interest, obtained by FDTD.}
    \label{fig:Figure1}
  \end{figure}

  The nanobeam was characterized using a cross-polarized confocal microscopy setup, as shown in Figure \ref{fig:Figure1}(a). In this arrangement, the light collected from the sample is of the orthogonal polarization from the light used to excite the sample, which allows the cavity emission to be observed without being overwhelmed by a strong background reflection \cite{article:eng07}. The nanobeam photonic crystal is fabricated from a 164 nm thick GaAs membrane with an embedded layer of InAs quantum dots using electron-beam lithography \cite{article:eng07}, and the cavity is formed by tapering the central holes and lattice constant \cite{loncar_highQnanobeam_APL2009, gong_nanobeamQDlaser_OpEx2010}. The photonic crystal is designed to have a lattice constant of $a=234$ nm and a hole size of $r=0.3a$; at the center of the beam the lattice constant is gradually decreased down to $a'=0.75a=175.5$ nm while decreasing the hole size to maintain $r'=0.3a'$. In addition, a dielectric region is left unpatterned at the center of the beam to allow a quantum dot to be located near the high-field region. A scanning electron micrograph (SEM) of the fabricated nanobeam cavity is shown in Figure \ref{fig:Figure1}(b). Based on this SEM, 3D finite-difference time-domain (FDTD) simulations were used to obtain the fundamental cavity resonance shown in Figure \ref{fig:Figure1}(c), depicting the $E_y$ field profile. These simulations predict a center wavelength of $\lambda=900$ nm, with a quality factor of $Q \approx 20,000$.

  \begin{figure}
    \includegraphics{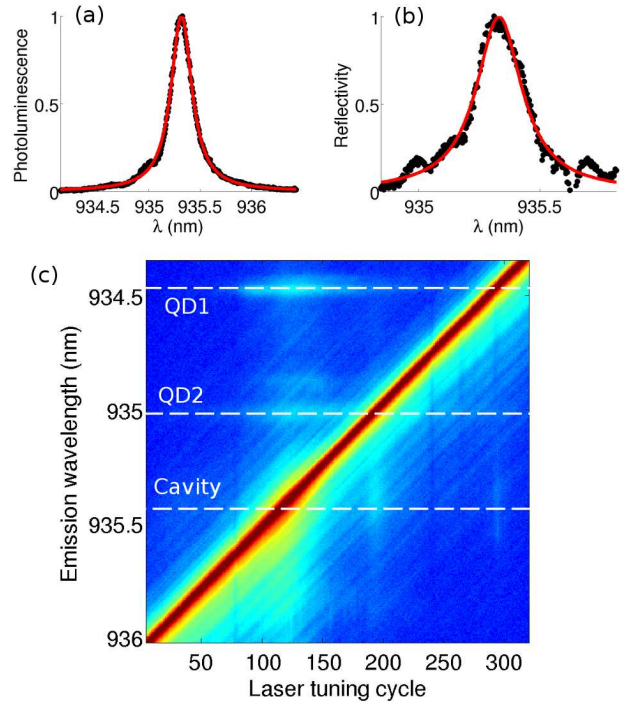}
    \caption{(Color online) Cavity emission from (a) QD photoluminescence and (b) laser reflectivity. (c) Laser scan showing off-resonant coupling between two quantum dots and the cavity.}
    \label{fig:Figure2}
  \end{figure}

  In order to characterize this cavity, a Ti:sapphire laser at 820 nm was used for above-band excitation of the quantum dots. The resulting photoluminescence caused by emission from the quantum dot layer then pumps the cavity, as shown in Figure \ref{fig:Figure2}(a). A Lorentzian fit yields a center wavelength of $\lambda=935.32$ nm and a quality factor of $Q=3,804$. Replacing the above-band source with a tunable, narrow-bandwidth CW laser tuned to the cavity wavelength yields the transmission measurement of the cavity resonance, as shown in Figure \ref{fig:Figure2}(b). As the laser is tuned across the cavity resonance, the cross-polarized reflectivity signal mimics a transmission study of the cavity. Here, a Lorentzian fit results in a center wavelength of $\lambda=935.33$ nm and a quality factor of $Q=4,058$. These measurements, while consistent with each other, do not match exactly with the results predicted by FDTD simulations. This discrepancy can be attributed to a thin layer of GaAs still present in the holes of the photonic crystal, which can be seen in the SEM of Fig.\ \ref{fig:Figure1}(b). This additional dielectric would have the observed effect of increasing the wavelength of the fundamental mode, as well as degrading $Q$ by reducing the index contrast and breaking the vertical symmetry of the photonic crystal.

  Scanning the excitation laser across the cavity resonance shows evidence of off-resonant coupling from the cavity to several nearby quantum dots, as shown in Figure \ref{fig:Figure2}(c). When the laser is on resonance with the cavity, emission is observed from quantum dots located at $\lambda = 934.5$ nm (QD1) and $\lambda = 935$ nm (QD2), although they are at higher energies relative to the excitation laser. We also observe the reverse effect, i.e., off-resonant coupling from each of these dots to the cavity: Figure \ref{fig:Figure2}(c) shows that when the laser is on resonance with either QD1 or QD2, emission is observed from the cavity.

  \begin{figure}
    \includegraphics{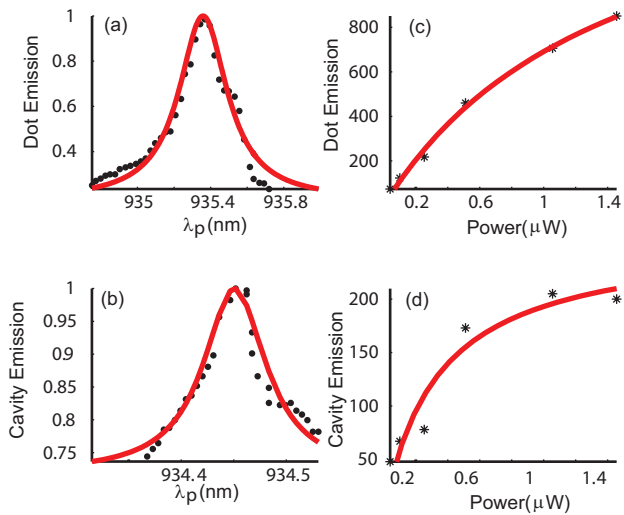}
    \caption{(Color online) (a) QD1 emission under cavity excitation, and (b) cavity emission under QD1 excitation, both as a function of probe laser wavelength $\lambda_p$. Power series showing saturation of both QD1 emission (c) and cavity emission (d) under the same excitation as in (a) and (b), respectively.}
    \label{fig:Figure3}
  \end{figure}

  In order to confirm that the observed emission is actually a result of off-resonant coupling, linewidth measurements can be extracted from the laser scan. Figure \ref{fig:Figure3}(a) shows the collected emission from QD1 (at $\lambda\sim 934.5$ nm) while the laser is scanned across the cavity resonance. The measured linewidth of $\sim0.3$ nm ($Q\approx3120$) and center wavelength of $\sim935.3$ nm matches the cavity resonance as measured earlier, even though the emission is collected from the quantum dot. The lower $Q$ measured in off-resonant coupling is consistent with previous results \cite{article:majumdar10}. Similarly, Figure \ref{fig:Figure3}(b) shows the collected emission from the cavity (at $\lambda\sim935.3$ nm) while the laser is scanned across the QD1 line, resulting in a measured linewidth of $\sim0.07$ nm and a center wavelength of $\sim934.45$ nm. As these figures show, both resonances display a nonsymmetric character not fully captured by the Lorentzian fit, as well as a small additional peak or shoulder that appears at a slightly longer wavelength ($\sim0.15$ nm longer in the case of dot emission under cavity excitation, and $\sim0.06$ nm longer in the case of cavity emission under dot excitation). This may simply be due to unintentional interference present in the optical path, or it could be a genuine property of off-resonant coupling in nanobeam cavities that deserves further exploration. The above measurements were conducted with a laser power of 253 nW before the OL, and at a temperature of 30K. However, off-resonant coupling from the cavity to the quantum dots was observed at temperatures as low as 15K, while coupling from a quantum dot to the cavity was observed as low as 25K.
  
  To ensure that this coupling is phonon-mediated and not due to generation of carriers by nonlinear optical processes \cite{rivoire_singlephoton_APL2011}, we performed a power dependent study of the cavity and the QD emission at a temperature of 30K, shown in Figure \ref{fig:Figure3}(c)-(d). As the laser power is increased, the emission from both the quantum dot (Fig.\ \ref{fig:Figure3}(c)) and the cavity (Fig.\ \ref{fig:Figure3}(d)) saturates. This saturation behavior originates from the two-level character of the QD, as has been modeled previously \cite{article:majumdar10, article:michler10, majumdar_phonon_11}.
  
  We have shown off-resonant coupling between a quantum dot and a nanobeam photonic crystal cavity, with photons being transferred from the optical cavity to the quantum dot, as well as the reverse. The demonstration of this process in a nanobeam geometry has interesting potential for future work. This is partly due to the broad versatility of nanobeams in many applications,  but is also a result of the fact that as a phonon-mediated process, off-resonant coupling is fundamentally dependent on the mechanical properties of the underlying structure. Pursuing control of nanobeam optomechanics, and thereby control of the off-resonant interaction of a quantum dot with an optical cavity, holds great promise for obtaining a better understanding of the role of phonons in solid state CQED.

  \begin{acknowledgments}
    The authors acknowledge financial support provided by the Office of Naval Research (PECASE Award), the Army Research Office, and the National Science Foundation. A.R. was supported by a Stanford Graduate Fellowship. J.V. would also like to acknowledge support from the Alexander von Humboldt Foundation. The QD material was provided by Pierre Petroff and Hyochul Kim at the University of California, Santa Barbara. This work was performed in part at the Stanford Nanofabrication Facility of NNIN, supported by the National Science Foundation.
  \end{acknowledgments}

\end{document}